\documentclass[runningheads]{svmult}

\usepackage{makeidx}   
\usepackage{graphicx}  
\usepackage{subeqnar}  
\usepackage{multicol}  
\usepackage{physprbb}  
\makeindex             

\begin{document}
\title*{Wide field photometry of the M104 globular cluster system}
%
%
%
%
%
\author{Alessia Moretti\inst{1,2}
\and Enrico V. Held\inst{1}
\and Luca Rizzi\inst{1,2}
\and Vincenzo Testa\inst{3}
\and Luciana Federici\inst{4}
\and Carla Cacciari\inst{4}}
\authorrunning{Alessia Moretti et al.}
%
%
\institute{INAF-Osservatorio Astronomico di Padova, 35122 Padova, Italy
\and Dipartimento di Astronomia,
     Universit\`a di Padova,
     I-35122 Padova, Italy 
\and INAF-Osservatorio Astronomico di Roma, 00040 Monte Porzio Catone, Italy
\and INAF-Osservatorio Astronomico di Bologna, 40127 Bologna, Italy}

\maketitle              

\begin{abstract}
We present preliminary results of a wide field study of the globular
cluster system of NGC\,4594, the Sombrero galaxy.  The galaxy was
observed in $B$, $V$, and $R$ using the Wide Field Imager on the ESO
2.2m telescope.  Using color and shape criteria to select a sample of
highly probable globular cluster candidates, we measured the radial
density profile of clusters out to 40$^\prime$ (100 Kpc) in the galaxy
halo.  The colors are consistent with the bimodal color distribution
observed in previous studies. The red cluster candidates show a clear
central concentration relative to the blue clusters. The population of
red clusters does not appear significantly flattened, thus indicating
that they are associated to the galaxy bulge rather than to the disk.
\end{abstract}

\section{Observations and reduction}
\begin{figure}
\centering
\includegraphics[width=1.\textwidth]{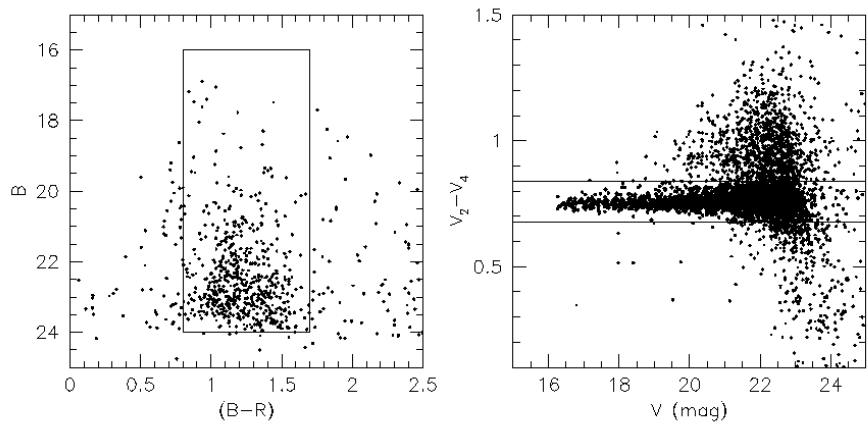}
\caption[]{ {\it Left:\ } the $B$ vs. $(B-R)$ color-magnitude diagram
of globular cluster candidates in the inner region of M\,104 ($ 1 < r
< 6$ arcmin). The rectangle indicates our selection of the most
probable candidates.  {\it Right:\ } the
concentration parameter used to discriminate point-like from diffuse
objects. Star-like objects are the sequence between the two horizontal
lines.}
\label{f_morettiF1}
\end{figure}

%
Wide field imaging of M\,104 was taken in April 2000 using the ESO/MPG
2.2m telescope at La Silla.  The typical exposure time was 35 min in
each of the $B$, $V$, and $R$ bands.
Pre-reduction was performed using the IRAF package MSCRED, while
distortion correction, image alignment onto a common sky reference
system, co-adding and mosaicing were obtained using a dedicated IRAF
package (Rizzi \& Held 2002, in prep.).
The combined images in each passband were then suitably processed to
subtract the galaxy's diffuse light. The ring median filtering in IRAF
was used to this purpose.
Finally, DAOPHOT aperture photometry was obtained on the sum images,
and a catalog was built of objects detected in all passbands at the
4$\sigma$ level.  From this master catalog, a sample of candidate
globular clusters was selected based on the object color and shape.
As a simple shape parameter we used the difference between two
aperture magnitudes with different radii of 2 and 4 pixels (the
typical size of stellar images being 4-5 pixels FWHM).  This simple
method proved to be quite effective in discriminating star-like images
from diffuse objects (i.e. background galaxies) 
(Fig.~\ref{f_morettiF1}, right panel).

The color-magnitude diagram of star-like objects around M\,104 clearly
shows the presence of a conspicuous globular cluster population (see
Fig.~\ref{f_morettiF1}, left panel).  The color and magnitude
selection outlined in Fig.~\ref{f_morettiF1} was used to 
define our final sample
of high-probability globular cluster candidates.  The adopted limits
were modeled on the 
properties of globular clusters in M\,31 \cite{B00}.  Using this
combined color and shape selection, contamination by foreground stars
and background galaxies in expected to be very low.


\section{Spatial distribution}

\begin{figure}
\centering
\includegraphics[width=.9\textwidth]{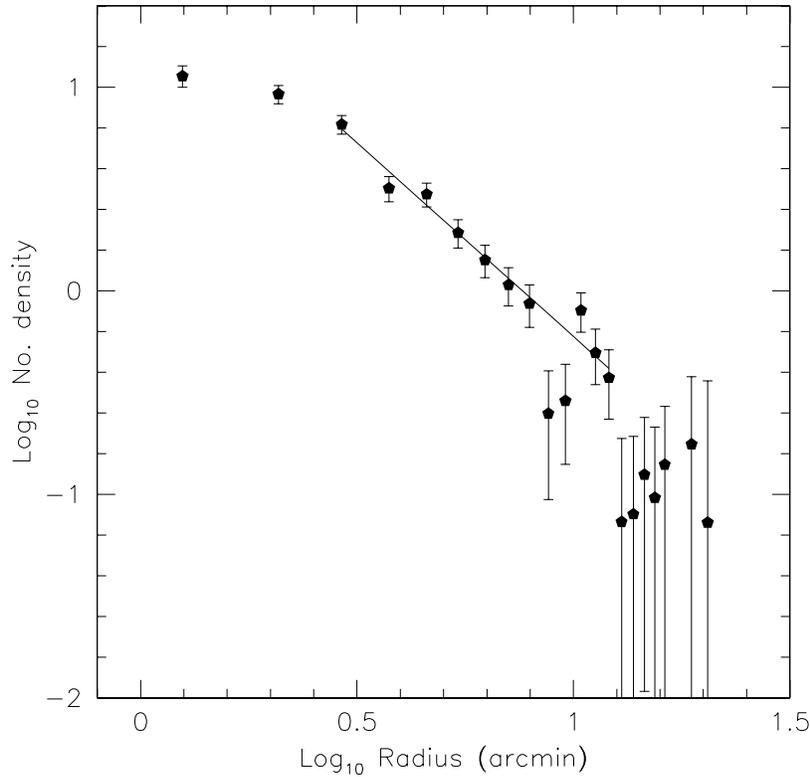}
\caption[]{The radial density profile of bona-fide globular cluster
candidates in M\,104.  A linear fit to the density profile between 2.5
and 12.5 arcmin is also shown, yielding a power-law exponent $-1.92$.}
\label{f_morettiF2}
\end{figure}

Wide field CCD imaging provides the possibility of detecting {\it
bona-fide} globular clusters out to large distances from the galaxy
center. 
Figure~\ref{f_morettiF2} shows the surface density profile of 
candidates within 15$^\prime$  (nearly 40 Kpc at the distance
of M\,104) from the galaxy center.  We have used the outer regions
(beyond 16.5 arcmin from the center) to estimate and statistically
subtract an approximate background level.
The cluster distribution appears to be more extended than the galaxy
light, confirming the result of the photographic study of W. Harris
and coll. \cite{HHH84}. There is also a hint of flattening in the
inner region, which is consistent with a two-slopes fit \cite{BH92}.

\section{Colors and metallicities}

\begin{figure}
\centering
\includegraphics[width=1.\textwidth]{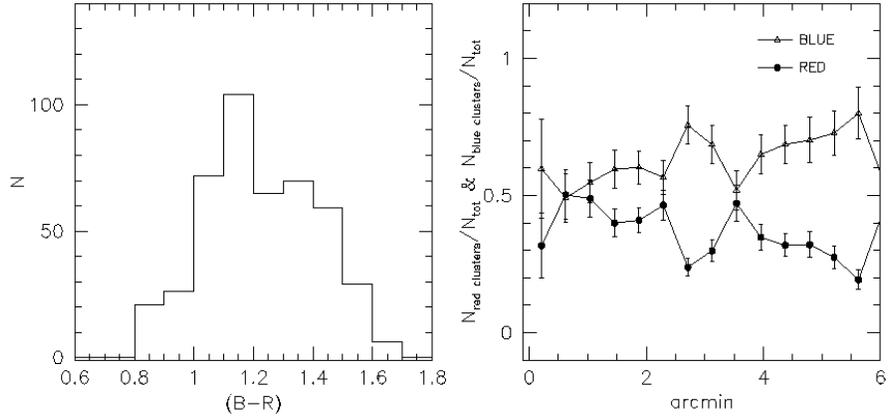}
\caption[]{{\it Left:\ } the $B-R$ color distribution of globular
clusters in the Sombrero galaxy.  {\it Right:\ } the radial
distribution of blue ($B-R < 1.3$, corresponding to [Fe/H]$ < -1$) and
red ($B-R \ge 1.3$) clusters, normalized to the total number of
cluster candidates in each bin.}
\label{f_morettiF3}
\end{figure}

The sample of {\it bona-fide} candidates was also used to derive the
color distribution of globular clusters in the halo of the Sombrero
galaxy.  Figure~\ref{f_morettiF3} suggests a bimodal metallicity
distribution, with peaks at $B-R =1.15$ and 1.35, corresponding to
[Fe/H]$ = -1.4$ and $-0.6$ by adopting the transformation of \cite{B00}.
The metallicities measured from colors and spectroscopy 
(\cite{L02}; Held et al., this volume) are in good agreement.
These results agree with those derived by Larsen and
coll. \cite{L01} from the $V-I$ colors of an HST inner cluster sample.
These metallicities are very similar to those of globular clusters in
our Galaxy and M\,31, where the blue component is associated to a
cluster sub-population which is old and metal-poor, while the red
component is metal-rich and typical of the bulge/disc population 
\cite{F00}.

\section{Radial gradients}

The red clusters turn out to be more centrally concentrated than the
blue ones, which confirms their association with the galaxy bulge
\cite{L01}.  The right panel of Fig.~\ref{f_morettiF3} shows the
radial distribution of the two sub-populations. Moving to the outer
regions the blue, presumably metal-poor, component becomes
predominant, and a ratio between red and blue clusters typical of
spiral galaxies is reached. 
Interestingly, the projected two-dimensional
distributions of both the blue and red sub-populations exhibit a
nearly spherical symmetry. As a results of our wide field multicolor
study, we can conclude that the red clusters are associated to the galaxy
bulge rather than to the disk.


%

\end{document}